\documentstyle[preprint,aps,floats,epsf]{revtex}

\begin{document}

\tightenlines 

\newcommand{\notE}{\ \hbox{{$E$}\kern-.60em\hbox{/}}}
\newcommand{\notp}{\ \hbox{{$p$}\kern-.43em\hbox{/}}}
\def\D0{\mbox{D\O}}


\includegraphics{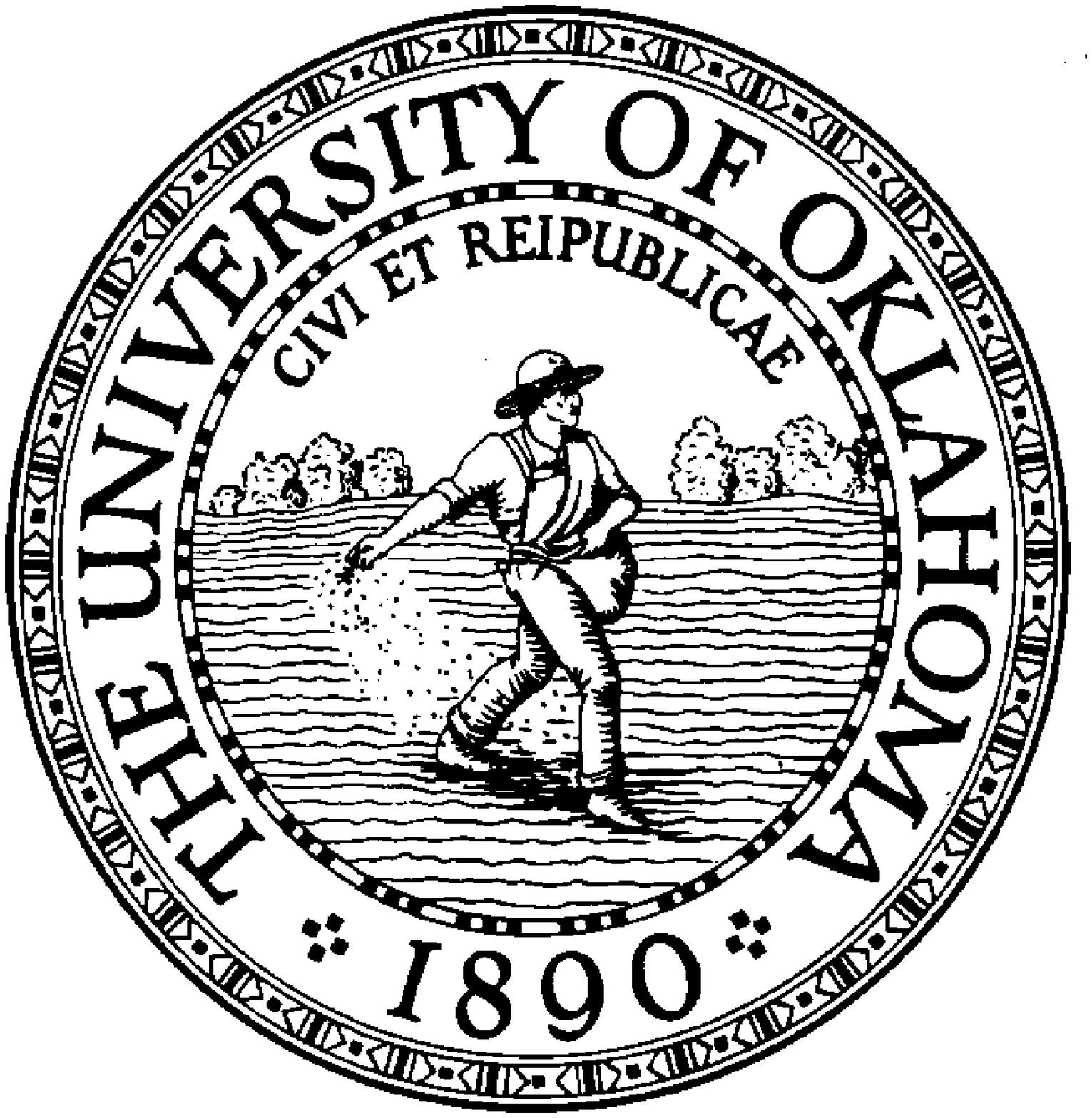}

\preprint{\font\fortssbx=cmssbx10 scaled \magstep2
\hbox to \hsize{
\hskip1.2in 
\hbox{\fortssbx The University of Oklahoma}
\hskip0.8in $\vcenter{
                      \hbox{\bf OKHEP-03-02}
                      \hbox{\bf hep-ph/0305028}
                      \hbox{May 2003}}$ }
}
 
\title{\vspace{0.5in}
Detecting a Higgs Pseudoscalar with a $Z$ Boson at the LHC}

\author{Chung Kao$^a$\footnote{E-mail address: Kao@physics.ou.edu}, 
Geoffrey Lovelace$^a$\footnote{Present address: Department of Physics,
California Institute of Technology, Pasadena, CA 91125.} and 
Lynne H. Orr$^b$\footnote{E-mail address: Orr@pas.rochester.edu}
}

\address{
$^a$Department of Physics and Astronomy, University of Oklahoma, 
Norman, OK 73019 \\
$^b$Department of Physics and Astronomy, University of Rochester, 
Rochester, NY 14627
}

\maketitle

\bigskip

\begin{abstract}

We have adopted two Higgs doublet models to study the production 
of a Higgs pseudoscalar ($A^0$) in association with a $Z$ gauge boson 
from gluon fusion ($gg \to ZA^0$) at the CERN Large Hadron Collider.
The prospects for the discovery of $ZA^0 \to \ell \bar{\ell} b\bar{b}$ 
are investigated with physics backgrounds and realistic cuts.
Promising results are found for $m_A \alt 260$ GeV in 
two Higgs doublet models when the heavier Higgs scalar ($H^0$) 
can decay into a $Z$ boson and a Higgs pseudoscalar ($A^0$).
Although the cross section of $gg \to ZA^0$ is usually small in the 
minimal supersymmetric standard model, it can be significantly enhanced 
in general two Higgs doublet models.
This discovery channel might provide an opportunity to search for 
a Higgs scalar and a Higgs pseudoscalar simultaneously at the LHC 
and could lead to new physics beyond the Standard Model 
and the minimal supersymmetric model.  

\end{abstract}

\pacs{pacs numbers: 12.15.ji, 13.85.qk, 14.80.er, 14.80.gt.}

\thispagestyle{empty}

\newpage

\section{INTRODUCTION}

The Standard Model has been very successful in explaining 
most experimental data to date, 
culminating in the discovery of the top quark \cite{topquark} and 
the evidence of the tau neutrino~\cite{nutau}. 
One of the most important experimental goals for Run~II of 
the Fermilab Tevatron and the CERN Large Hadron Collider (LHC) 
is the experimental investigation of the mechanism behind 
electroweak symmetry breaking---the discovery of the Higgs bosons 
or the proof of their non-existence, 
and the search for higher symmetries beyond the Standard Model.

In the Standard Model (SM), the Higgs mechanism requires 
only one Higgs doublet to generate masses for fermions and gauge bosons. 
It leads to the appearance of a neutral CP-even Higgs scalar 
after electroweak symmetry breaking (EWSB). 
The LEP2 experiments have established a lower bound of 114.4~GeV \cite{LEP2a} 
for the SM Higgs boson mass at the 95\% confidence level.

A general two Higgs doublet model (2HDM) \cite{guide} has Higgs doublets 
with the vacuum expectation values $v_1$ and $v_2$ 
that are needed to give masses to both down-type and up-type quarks 
as well as leptons and gauge bosons. 
There are five physical Higgs bosons: a pair of singly charged Higgs bosons 
$H^{\pm}$, two neutral CP-even scalars $H^0$ (heavier) and $h^0$ (lighter), 
and a neutral CP-odd pseudoscalar $A^0$. 
At the tree level, the couplings of the Higgs bosons to fermions 
and gauge bosons are determined by six independent parameters: 
the four Higgs boson masses, the ratio of vacuum expectation values 
$\tan\beta \equiv v_2/v_1$, and a mixing angle $\alpha_H$ between the 
weak and mass eigenstates of the neutral scalars.

The minimal supersymmetric standard model (MSSM) \cite{MSSM} requires 
two Higgs doublets to generate masses for fermions and gauge bosons 
and to cancel triangle anomalies associated with the fermionic partners 
of the Higgs bosons. 
At the tree level, the Higgs sector has only two free parameters 
that are commonly selected to be $m_A$ and $\tan\beta$. 
The mixing angle $\alpha_H$ between the neutral scalars is often chosen to be 
negative ($-\pi/2 \leq \alpha_H \leq  0$).
The LEP2 collaborations have set a lower bound of 91~GeV and 91.9 GeV 
\cite{LEP2b} 
for the $m_h$ and the $m_A$, respectively. 

Extensive studies have been made for the detection of the heavier 
MSSM Higgs bosons ($H^0$ and $A^0$) at the CERN LHC 
\cite{HGG,Neutral,Kunszt,HXX2,AZh,Nikita,Sally,CMS,ATLAS1,ATLAS2}. 
For $\tan\beta \alt 5$, 
$A^0 \to \gamma\gamma$, $H^0 \to ZZ$ or $ZZ^*\to 4l$,  
and $A^0,H^0 \to t\bar{t}$ are possible discovery channels.
The detection modes $A^0 \to Zh^0 \to l^+l^- \tau\bar{\tau}$ \cite{AZh} 
or $l^+l^- b\bar{b}$ \cite{AZh,CMS,ATLAS2} 
and $H^0 \to h^0 h^0 \to b\bar{b} \gamma\gamma$ \cite{ATLAS2} 
may be promising channels for simultaneous discovery of 
two Higgs bosons in the MSSM. 
For large values of $\tan\beta$, 
the $\tau\bar{\tau}$ decay mode \cite{Kunszt,CMS,ATLAS1,ATLAS2} 
and the muon pair decay mode \cite{Nikita,Sally,CMS,ATLAS2} 
are promising discovery channels for the $A^0$ and the $H^0$. 
In some regions of parameter space, the rates for Higgs boson
decays to neutralinos ($H^0,A^0 \to \chi^0_2 \chi^0_2$) are dominant and 
they might open up new promising modes for Higgs detection \cite{HXX2}.

In two Higgs doublet models, there are two complementary channels 
to search for a Higgs scalar and a Higgs pseudoscalar simultaneously: 
(i) $A^0 \to Zh^0$ \cite{AZh,CMS,ATLAS2} 
with a coupling proportional to $\cos(\beta-\alpha_H)$ and 
(ii) $H^0 \to ZA^0$ 
with a coupling proportional to $\sin(\beta-\alpha_H)$. 
At the LHC with high energy, the fraction $x$ of the parton momentum 
to the initial proton momentum can become small and greatly enhance 
the gluon-gluon luminosity. 
Therefore, gluon fusion can be a significant source of producing 
a Higgs pseudoscalar ($A^0$) and a $Z$ boson ($gg \to ZA^0$) 
via triangle and box diagrams with the third generation quarks 
\cite{ggza,Yin:2002sq}.

In this article, we present the prospects of discovering 
a Higgs pseudoscalar ($A^0$) associated with a $Z$ boson produced at the LHC 
followed by $Z \to \ell\bar{\ell}$ and $A^0 \to b\bar{b}$.
We evaluate the cross section for the Higgs signal and 
the complete SM background $pp \to \ell\bar{\ell}  b\bar{b} +X$ 
with realistic cuts and study the discovery potential at the LHC.
The production cross sections of $ZA^0$ at the LHC 
in a two Higgs doublet model and the MSSM are discussed in Section II.
The dominant physics backgrounds from production of 
$\ell\bar{\ell}b\bar{b}$ and $W^+ W^- b\bar{b}$ are presented in Section III.
The observability of $ZA^0 \to \ell\bar{\ell}b\bar{b}$ is discussed 
in Section IV.
Conclusions are drawn in Section V.

\section{THE PRODUCTION CROSS SECTIONS}

We calculate the cross section for $pp \to ZA^0 +X$ via $gg \to ZA^0$ 
within the framework of two Higgs doublet models (2HDM) with Model II 
of the Yukawa interactions for Higgs bosons and fermions \cite{Model2}. 
In Model II of 2HDMs and the MSSM, 
one Higgs doublet ($\phi_1$) couples to down-type quarks and charged leptons 
while another doublet ($\phi_2$) couples to up-type quarks and neutrinos.

The production of $ZA^0$ from gluon fusion involves triangle and box diagrams 
with loop integrals that are expressible in terms of the Spence functions 
\cite{Tini1,Tini2}. 
In our analysis, the one loop integrals were evaluated numerically with 
a FORTRAN code developed for one--loop diagrams \cite{LOOP}.
The parton distribution functions of CTEQ6L1 \cite{CTEQ6} 
are employed to evaluate the cross section for 
$pp \to ZA^0 \to \ell\bar{\ell} b\bar{b} +X$ 
with the Higgs production cross section $\sigma(pp \to ZA^0 +X)$ 
multiplied by the branching fractions of $Z \to \ell\bar{\ell}$ 
and $A^0 \to b\bar{b}$.

In Figure 1, we present the cross section 
$pp \to ZA^0 \to \ell\bar{\ell} b\bar{b} +X$ as a function of $\tan\beta$ 
in (a) the minimal supersymmetric model and 
(b) a two Higgs doublet model with $M_H = m_A + 100$ GeV, $m_h = 120$ GeV, 
and $\alpha_H = -\pi/4$. 
It is clear that the cross section in a 2HDM can be significantly larger 
than that in the the MSSM. 
Since $M_H$, $m_h$ and $\alpha_H$ are free parameters in a 2HDM, 
the $H^0$ can decay into $ZA^0$ with $m_H > m_A +M_Z$. 
In the MSSM with $\tan\beta \agt 10$,
$m_A$ and $m_h$ are very close to each other for $m_A \alt$ 125 GeV,  
while $m_A$ and $m_H$ are almost degenerate when $m_A \agt$ 125 GeV 
\cite{Nikita}. 
Therefore, the decay $H^0 \to ZA^0$ is kinematically inaccessible.

The cross section for $pp \to Z A^0 +X \to \ell\bar{\ell}b\bar{b} +X$ 
is shown in Figure 2 as a function of the Higgs scalar mixing angle 
$\alpha_H$,  in a two Higgs doublet model for $\tan\beta =$ 2, 10, and 40, and 
(a) $m_A = 150$ GeV and (b) $m_A = 400$ GeV.
Also shown are the cross sections in the MSSM 
for $\tan\beta = 2$ (diamond), 10 (square), and 40 (circle).
For $\alpha_H < 0$, the cross section in a 2HDM is significantly larger 
than that in the MSSM except when $\alpha_H \sim -\pi/2$.
For $m_A >$ 250 GeV and $\tan\beta \alt 7$, 
the branching fraction of $A^0 \to b\bar{b}$ 
is greatly suppressed when the Higgs pseudoscalar decays dominantly  
into $t\bar{t}$ with one of the top quarks being virtual.

We note that there are dips in Figures 1 and 2. 
For some values of $\tan\beta$ and $\alpha_H$, 
the cross section of $ZA^0$ from gluon fusion is highly suppressed 
by the destructive interference between the triangle
and the box diagrams as well as the negative interference between
the top quark and the bottom quark loops, 
especially when they are comparable \cite{ggza}.

\section{THE PHYSICS BACKGROUND}
 
The dominant physics backgrounds to the final state of 
$ZA^0 \to \ell\bar{\ell}b\bar{b}$ come from $gg \to \ell\bar{\ell}b\bar{b}$ 
and $q\bar{q} \to \ell\bar{\ell}b\bar{b}, \ell = e$ or $\mu$. 
The background from $pp \to W^+W^- b\bar{b} +X$ 
(including $pp \to t\bar{t} +X$) 
followed by the decays of $W^\pm \to \ell^\pm \nu_\ell$, can be effectively 
reduced with cuts on the invariant mass of lepton pairs and 
the missing transverse energy. 
We have also considered backgrounds from 
$pp \to \ell\bar{\ell} gb +X$, $pp \to \ell\bar{\ell} g\bar{b} +X$, 
$pp \to \ell\bar{\ell} gq +X$, $pp \to \ell\bar{\ell} g\bar{q} +X$, and 
$pp \to \ell\bar{\ell} jj +X$, 
where $q = u, d, s$, or $c$ and $j = g, q$ or $\bar{q}$.

Our acceptance cuts and efficiencies of $b$-tagging and mistagging 
are similar to those of the ATLAS collaboration \cite{ATLAS2}. 
In each event, two isolated leptons are required to have 
$p_T(\ell) > 15$ GeV and $|\eta(\ell)| < 2.5$. 
For an integrated luminosity ($L$) of 30 fb$^{-1}$, 
we require $p_T(b,j) > 15$ GeV and $|\eta(b,j)| < 2.5$. 
The $b$-tagging efficiency ($\epsilon_b$) is taken to be $60\%$, 
the probability that a $c$-jet is mistagged as a $b$-jet ($\epsilon_c$)
is $10\%$, and 
the probability that any other jet is mistagged as a $b$-jet ($\epsilon_j$)
is taken to be $1\%$.
Furthermore, we require the invariant mass of the opposite sign pair 
of leptons to be within 10 GeV of $M_Z$, 
that is $|M_{\ell\bar{\ell}}-M_Z| \le 10$~GeV. 

For a higher integrated luminosity of 300 fb$^{-1}$, 
we require the same acceptance cuts as those for $L =$ 30 fb$^{-1}$, 
except $p_T(\ell) > $ 25 GeV and $p_T(b,j) > 30$ GeV. 
The $b$-tagging efficiency ($\epsilon_b$) is taken to be $50\%$, 
and the probability that a $c$-jet is mistagged as a $b$-jet ($\epsilon_c$)
is $14\%$. 

In addition, we require that the missing transverse energy ($\notE_T$) 
in each event should be less than 20 GeV for $L = 30$ fb$^{-1}$ 
and less than 40 GeV for $L = 300$ fb$^{-1}$. This cut effectively 
reduces the background from $pp \to W^+W^- b\bar{b} +X$ which receives 
the major contribution from both real and virtual top quarks 
$pp \to t^* \bar{t}^* +X$.

We have employed the programs MADGRAPH \cite{madgraph}
and HELAS \cite{helas} to evaluate
the background cross sections of 
$pp \to \ell\bar{\ell}b\bar{b} +X, \ell\bar{\ell} jj +X$,  
and $W^+W^-b\bar{b} +X$.
The $p_T(b,j)$ cut is effective in removing 
most of the SM background, while most $b$ and $\bar{b}$ 
from the Higgs decays pass the $p_T$ cut.

In Figure 3, we present distributions for the transverse momenta ($p_T$)
of leptons and bottom quarks
for $pp \to ZA^0 \to \ell\bar{\ell} b\bar{b} +X$
and for the SM background of $pp \to \ell\bar{\ell} b\bar{b} +X$ via
$gg \to \ell\bar{\ell} b\bar{b}$ and $q\bar{q} \to \ell\bar{\ell} b\bar{b}$. 
It is clear that the $p_T$ cuts on leptons and bottom quarks are effective 
in removing most of the SM background, while most leptons from the $Z$ decays 
and most $b$'s from the Higgs decays survive the $p_T$ cuts.
 
\section{THE DISCOVERY POTENTIAL AT THE LHC}

To study the discovery potential of 
$pp \to Z A^0 \to \ell\bar{\ell} b\bar{b} +X$ at the LHC, 
we calculate the background from the SM processes of 
$pp \to \ell\bar{\ell} b\bar{b} +X$ in the mass window of
$m_A \pm \Delta M_{b\bar{b}}$ with $\Delta M_{b\bar{b}} = 22$ GeV.

We consider the Higgs signal to be observable 
if the $N \sigma$ lower limit on the signal plus background is larger than 
the corresponding upper limit on the background \cite{HGG,Brown}, namely,
\begin{eqnarray}
L (\sigma_S+\sigma_B) - N\sqrt{ L(\sigma_S+\sigma_B) } > 
L \sigma_B +N \sqrt{ L\sigma_B }
\end{eqnarray}
which corresponds to
\begin{eqnarray}
\sigma_S > \frac{N^2}{L} \left[ 1+2\sqrt{L\sigma_B}/N \right] \, .
\end{eqnarray}
Here $L$ is the integrated luminosity, 
$\sigma_S$ is the cross section of the Higgs signal, 
and $\sigma_B$ is the background cross section 
within a bin of width $\pm \Delta M_{b\bar{b}}$ centered at $m_A$. 
In this convention, $N = 2.5$ corresponds to a 5$\sigma$ signal.

We show the cross section with acceptance cuts in Figure 4 for 
$pp \to ZA^0 \to \ell\bar{\ell} b\bar{b} +X$ 
in a 2HDM with $M_H = m_A + 100$ GeV, $m_h = 120$ GeV, 
and $\alpha_H = -\pi/4$ for an integrated luminosity of 30 fb$^{-1}$ 
and a higher luminosity of 300 fb$^{-1}$.
The curves for the 5$\sigma$ and 3$\sigma$ cross sections 
for the $ZA^0$ signal are also presented.

With a luminosity of 30 fb$^{-1}$, it is possible to establish a 
3$\sigma$ signal of $ZA^0 \to \ell\bar{\ell} b\bar{b}$ 
for $m_A \alt 200$ GeV and $\tan\beta$ close to 2. 
At a higher luminosity of 300 fb$^{-1}$ the discovery potential 
of this channel is greatly improved for $m_A \alt 250$ GeV.

In Tables I and II, we present event rates after acceptance cuts 
for the signal ($N_S$) and the background ($N_B$) as well as 
the ratio of signal to background $N_S/N_B$ and $N_S/\sqrt{N_B}$ 
in a two Higgs doublet model with $\tan\beta = 2$, $\alpha_H = -\pi/4$. 
It is clear that our acceptance cuts can effectively reduce  
the physics background so that the ratio $N_S/N_B$  is larger than 
$5\%$ for $M_A \alt 250$ GeV. These data demonstrate that our 
statistical argument in Eq. (2) is valid in determining 
the signal observability.

\begin{table}[h]
\caption{
Event rates after acceptance cuts for the signal ($N_S$) 
and the background ($N_B$) as well as 
the ratio of signal to background $N_S/N_B$ and $N_S/\sqrt{N_B}$ 
in a two Higgs doublet model with $\tan\beta = 2$, $\alpha_H = -\pi/4$, 
and $m_H = m_A + 100$ GeV for an integrated luminosity of 30 fb$^{-1}$.}
\centering\unskip\smallskip
\tabcolsep=1.5em
\begin{tabular}{c|cccc}
\hline
$m_A$ (GeV) & $N_S$ & $N_B$ & $N_S/N_B$ & $N_S/\sqrt{N_B}$ \\
\hline
100  &  690 &  9939  & 1/14  &  6.9 \\
150  &  252 &  4218  & 1/17  &  3.9 \\
200  &  387 &  1950  & 1/5   &  8.8 \\
250  &   50 &   978  & 1/19  &  1.6 \\
300  &    6 &   525  & 1/88  &  0.3 \\
350  &    1 &   299  & 1/360 &  0.05 \\
\hline
\end{tabular}
\end{table}

\begin{table}[h]
\caption{
The same as in Table I, 
except that the integrated luminosity is 300 fb$^{-1}$.}
\centering\unskip\smallskip
\tabcolsep=1.5em
\begin{tabular}{c|cccc}
\hline
$m_A$ (GeV) & $N_S$ & $N_B$ & $N_S/N_B$ & $N_S/\sqrt{N_B}$ \\
\hline
100  & 2821 & 22527 & 1/8   & 19   \\
150  & 1234 & 13461 & 1/11  & 11   \\
200  & 2062 &  7344 & 1/3.6 & 24   \\
250  &  276 &  4152 & 1/15  &  4.3 \\
300  &   33 &  2435 & 1/73  &  0.7 \\
350  &    5 &  1483 & 1/310 &  0.1 \\
\hline
\end{tabular}
\end{table}

\section{CONCLUSIONS}
 
We have found promising results 
for $pp \to ZA^0 \to \ell\bar{\ell}b\bar{b} +X$ 
via $gg \to ZA^0$ in two Higgs doublet models 
at the LHC with $L = 300$ fb$^{-1}$ for $m_A \alt 260$ GeV, 
$M_H = m_A +100$ GeV and $\tan\beta \sim 2$. 
The physics background in the Standard Model can be greatly reduced 
with suitable acceptance cuts.

In the MSSM with $m_A \agt 125$ GeV, $m_A \sim m_H$, and 
the production cross section of $gg \to ZA^0$ is usually small.  
The production rate of $ZA^0$ from gluon fusion at the LHC is
highly suppressed owing to the destructive interference between the triangle
and the box diagrams as well as the negative interference between
the top quark and the bottom quark loops, 
especially when they are comparable \cite{ggza}.

In a two Higgs doublet model, the cross section of $gg \to ZA^0$ 
can be greatly enhanced when the heavier scalar can decay into 
the pseudoscalar and a $Z$ boson. 
If we take $m_H \sim m_A$ in a 2HDM, the Higgs signal will be reduced 
to the level of the MSSM. 
This discovery channel might lead to new physics beyond 
the Standard Model and the minimal supersymmetric model. 
Furthermore, we might be able to discover two Higgs bosons 
simultaneously if the heavier Higgs scalar ($H^0$) can decay into 
a $Z$ boson and a Higgs pseudoscalar ($A^0$).

\section*{Acknowledgments}

This research was supported in part by the U.S. Department of Energy 
under grants 
No.~DE-FG03-98ER41066, 
No.~DE-FG02-03ER46040, 
and 
No.~DE-FG02-91ER40685.
 



\begin{figure}
\centering\leavevmode
\epsfxsize=6in\epsffile{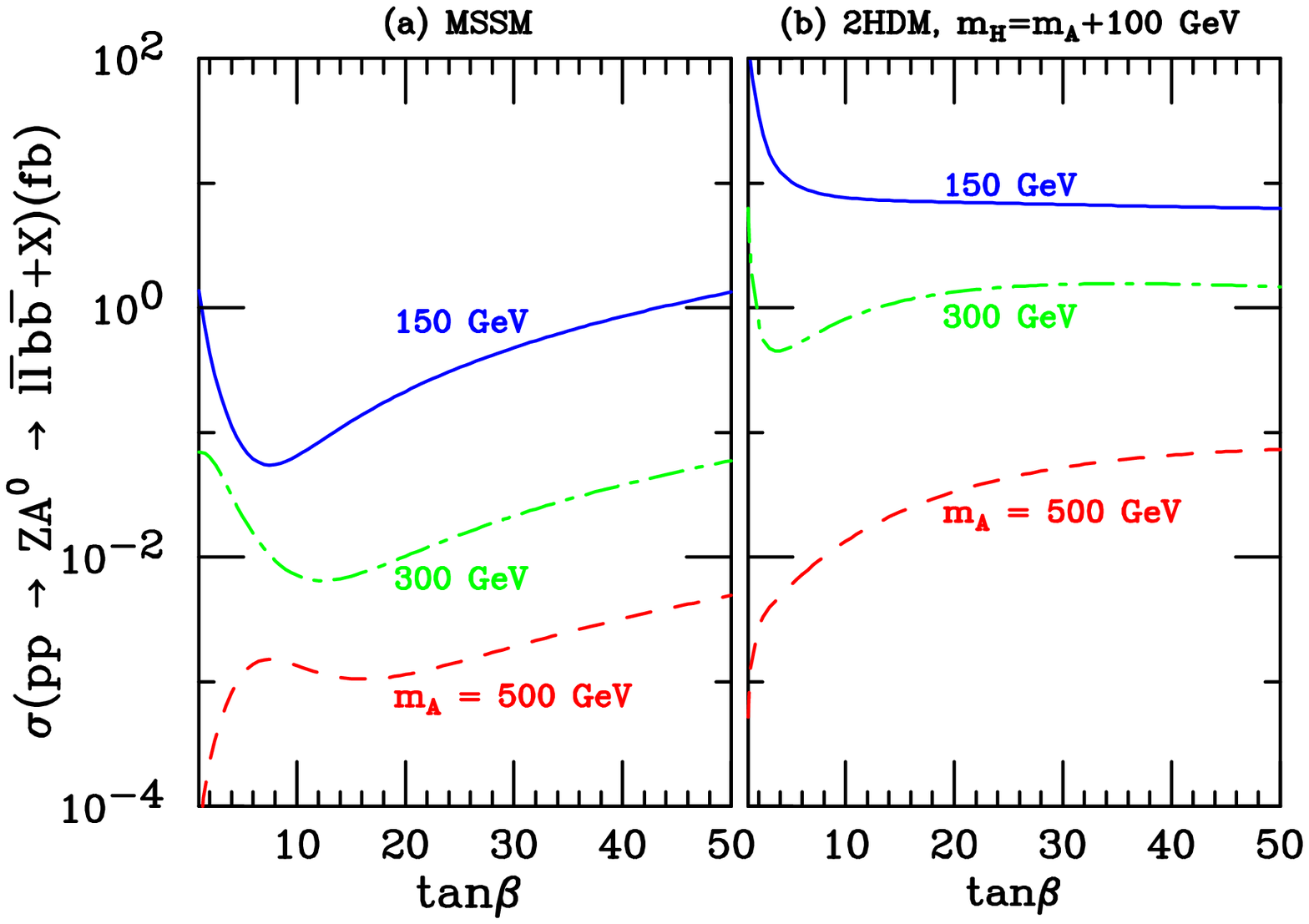}

\caption[]{
The cross section for $pp \to Z A^0 +X \to \ell\bar{\ell}b\bar{b} +X$ 
in fb without cuts at $\sqrt{s} = 14$ TeV, as a function of $\tan\beta$, 
for $m_A = 150, 300$, and $500$ GeV, 
in (a) the MSSM with $m_{\tilde{q}} = m_{\tilde{g}} = \mu = 1$ TeV, 
as well as 
in (b) a two Higgs doublet model with $M_h = 120$ GeV, $M_H = m_A +100$ GeV 
and the Higgs mixing angle $\alpha_H = -\pi/4$.
\label{fig:tanb}
}\end{figure}


\begin{figure}
\centering\leavevmode
\epsfxsize=6in\epsffile{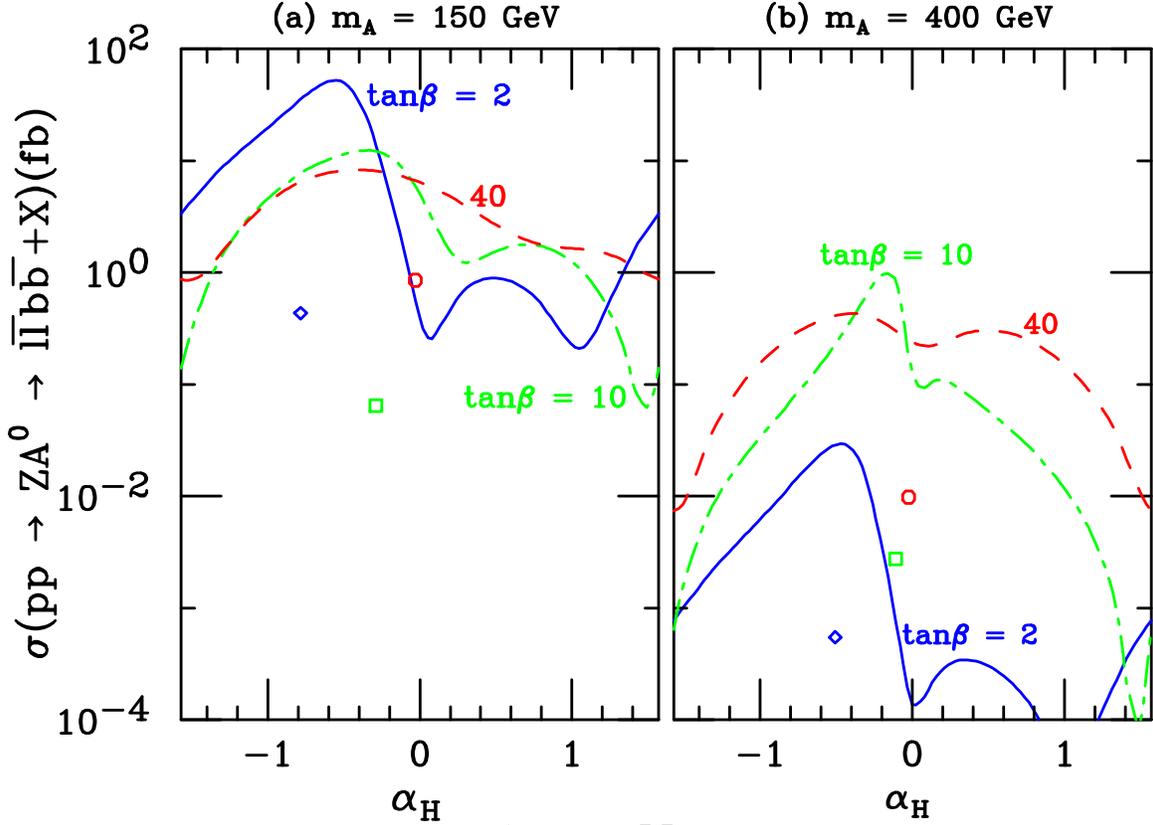}

\caption[]{
The cross section for $pp \to Z A^0 +X \to \ell\bar{\ell}b\bar{b} +X$ 
in fb without cuts at $\sqrt{s} = 14$ TeV, 
as a function of the Higgs scalar mixing angle $\alpha_H$,  
in a two Higgs doublet model with $M_h = 120$ GeV, $M_H = m_A +100$ GeV 
for $\tan\beta =$ 2, 10, and 40, and 
(a) $m_A = 150$ GeV and (b) $m_A = 400$ GeV.
Also shown are the cross sections in the MSSM 
for $\tan\beta = 2$ (diamond), 10 (square), and 40 (circle).
\label{fig:alpha_H} 
}\end{figure}


\begin{figure}
\centering\leavevmode
\epsfxsize=6in\epsffile{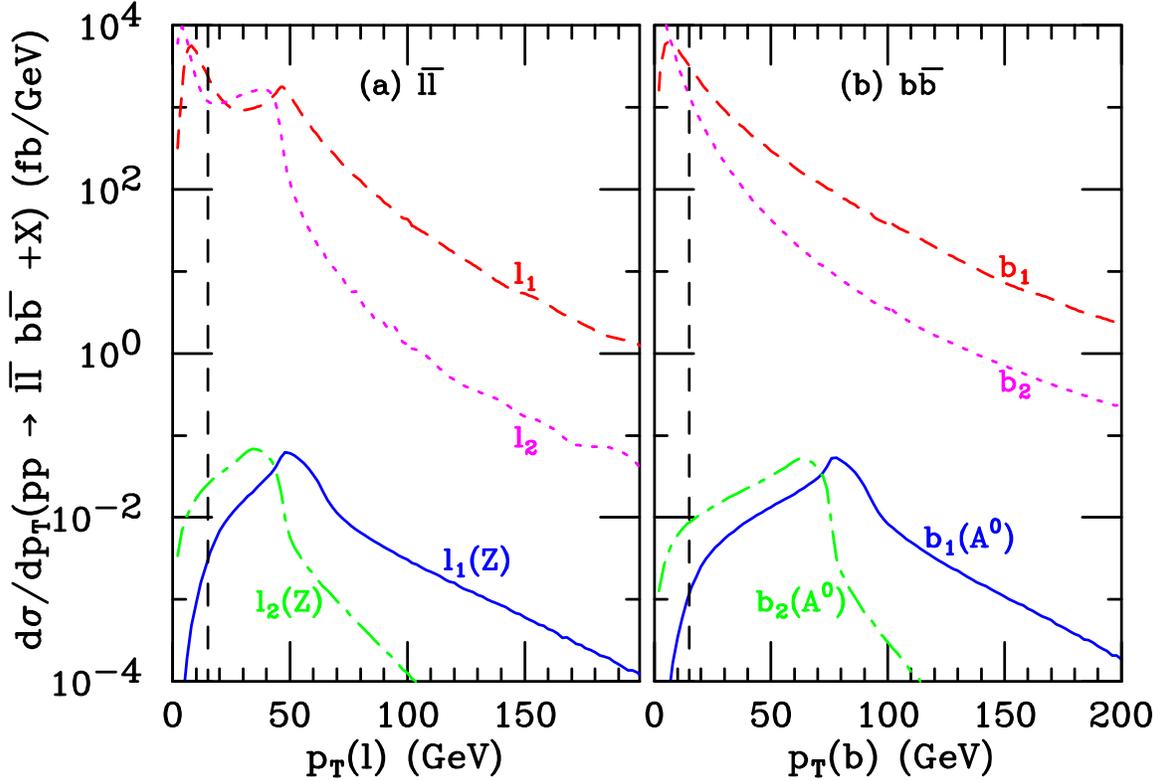}

\caption[]{
The $p_T$ distributions of leptons and bottom quarks in fb/GeV 
for $pp \to Z A^0 \to \ell\bar{\ell} b\bar{b} +X$ (solid and dot-dashed) 
in a two Higgs doublet model with 
$m_A = 150$ GeV, $m_h = 120$ GeV, $m_H = m_A +100$ GeV, 
the Higgs scalar mixing angle $\alpha = -\pi/4$, 
and $\tan\beta = 40$.
We have chosen $\ell_1$ or $b_1$ to be the lepton or the b quark with higher 
transverse momentum.
Also shown is the contribution from the SM background of 
$pp \to \ell\bar{\ell} b\bar{b} +X$ (dashed and dotted) via
$gg \to \ell\bar{\ell} b\bar{b}$ and $q\bar{q} \to \ell\bar{\ell} b\bar{b}$.
The $p_T$ cuts remove events to the left of the vertical dashed line.
\label{fig:PTlb}
}\end{figure}


\begin{figure}
\centering\leavevmode
\epsfxsize=6in\epsffile{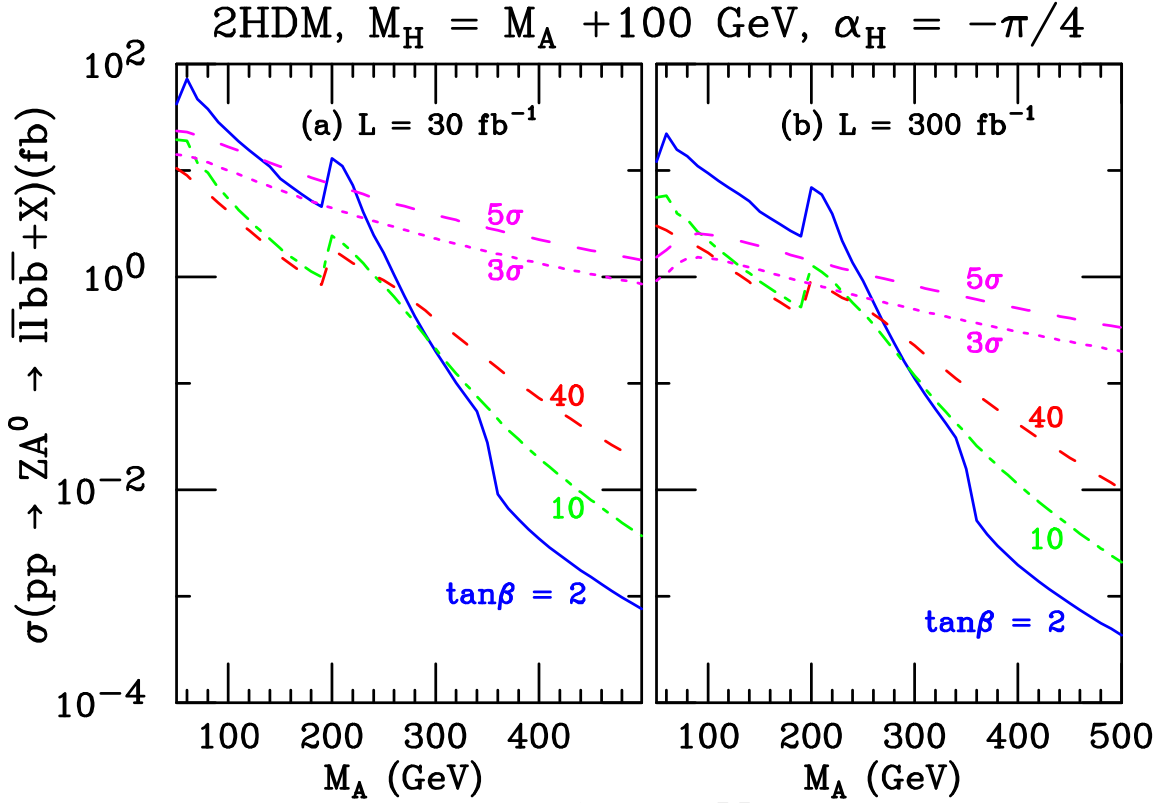}

\caption[]{
The cross section in fb of $pp \to Z A^0 +X \to \ell\bar{\ell} b\bar{b} +X$ 
versus $m_A$, at $\sqrt{s} = 14$ TeV, 
in a two Higgs doublet model with $m_h = 120$ GeV, $m_H = m_A +100$ GeV 
and the Higgs scalar mixing angle $\alpha_H = -\pi/4$, 
for $\tan\beta = 2$ (solid), 10 (dot-dashed), and 40 (dashed). 
Also shown are the 5$\sigma$ (dashed) and 3$\sigma$ (dotted)  
cross sections for the $ZA^0$ signal required 
for an integrated luminosity~($L$) of (a) 30 fb$^{-1}$ and (b) 300 fb$^{-1}$. 
We have applied the acceptance cuts as well as 
the tagging and mistagging efficiencies described in the text.
\label{fig:discovery}
}\end{figure}

\end{document}